\documentclass[12pt]{article}
\usepackage{graphicx}
\def\baselinestretch{1.5}
\def\gsim{\stackrel{>}{\sim}}
\def\lsim{\stackrel{<}{\sim}}
\def\beq{\begin{equation}}
\def\eeq{\end{equation}}
\def\ba{\begin{array}}
\def\ea{\end{array}}
\def\ol{\overline}

\def\req#1{(\ref{#1})}
\def\eq#1{Eq.~(\ref{#1})}
\def\pri{^{\, \prime}}

\def\deg{\ifmmode{^{\circ}}\else ${^{\circ}}$\fi}

\def\itm#1{\item[$(#1)$]}
\def\gsim{\,\raisebox{-0.13cm}{$\stackrel{\textstyle>}{\textstyle\sim}$}\,}
\def\lsim{\,\raisebox{-0.13cm}{$\stackrel{\textstyle<}{\textstyle\sim}$}\,}
\def\bi{\begin{itemize}}
\def\ei{\end{itemize}}
\def\ed{\end{document}}
\def\be{\begin{equation}}
\def\ee{\end{equation}}
\def\bea{\begin{eqnarray}}
\def\eea{\end{eqnarray}}
\def\beas{\begin{eqnarray*}}
\def\eeas{\end{eqnarray*}}
\def\req#1{(\ref{eq:#1})}
\def\eq#1{Eq.~(\ref{eq:#1})}
\def\labeq#1{\label{eq:#1}}

   \def\half{\frac{1}{2}}

   \def\egzk{E_{\rm GZK}}
   \def\tentwenty{10^{20}}
   
\def\dmpc{D_{\rm Mpc}}
\def\lammpc{\lambda_{\rm Mpc}}
   
\begin{document}
   \begin{flushright}
   preprint VAND--TH--00--7, NUB-3210/00-Th\\
   \end{flushright}
   \begin{centering}
  {\Large Clustering in Highest Energy Cosmic Rays:\\
   Physics or Statistics?}\\
   \vspace{1cm}
   {\large Haim Goldberg}$^1$ {\large and Thomas J. Weiler}$^2$\\
   \vspace{0.5cm}
   {\it $^1$ Department of Physics, Northeastern University, Boston, MA
   02115\\
  $^2$ Department of Physics \& Astronomy, Vanderbilt University,\\
   Nashville, TN 37235}\\
   \end{centering}
   \vspace{1cm}
   \begin{abstract}
  Directional clustering can be expected in cosmic ray observations due to purely
  statistical fluctuations for sources distributed randomly in the sky. We
develop an analytic approach to estimate the probability of random cluster
configurations, and use these results to study the strong potential of
the HiRes, Auger, Telescope
Array and EUSO/OWL/AirWatch facilities for
deciding whether any observed clustering is most likely due to non-random sources.
\\
   \end{abstract}
   \section*{Introduction}
   An unsolved astrophysical
mystery is the origin and nature of the extreme energy
   cosmic ray primaries (EECRs) responsible for the observed events
at highest energies, $\sim\tentwenty$~eV~\cite{EECRreviews}.
About twenty events above $\egzk$
   have been observed by five different experiments~\cite{Xpts}.
   The longitudinal profile for one of these events
   (the highest energy Fly's Eye event at $3\times 10^{20}$~eV)
   is available; it favors a nucleon or nuclear primary over a
photon primary \cite{GS00}.
   The origin of these events is a mystery,
for there are no visible source candiates
   within 50 Mpc except possibly M87, a radio-loud AGN at $\sim 20$~Mpc.
   Since the observed events display a large-scale isotropy, many sources
   rather than one source seems to be required. The nature of the primary is
   a mystery, because interactions with the 2.73K cosmic microwave background
(CMB) renders the Universe opaque to nucleons at these energies, and double
pair production on the cosmic radio background (CRB) renders the Universe
opaque to photons at these energies. The theoretical prediction of the end of
transparency for nucleons at $\egzk\sim 5\times \tentwenty$~eV is the famous
``GZK cutoff'' \cite{gzk}.

Models for the origin and nature of the primaries may be put into two broad
categories.
   %
   %
In the first category are postulated sources of protons and photons ``locally''
distributed within 50 to 100 Mpc. For these models, the propagation problem is
mitigated. However, the source problem is aggravated. In traversing a distance
$D$, a charged particle interacting with magnetic domains having coherence
length $\lambda$ will bend through an energy-dependent angle\footnote {On
average, half of the interactions of a super-GZK nucleon with the CMB change
the isospin.  At energies for which $c\tau$ of the neutron is small compared to
the interaction mfp of $\sim 6$~ Mpc, the neutron decays back to a proton with
negligible energy loss and the bending-angle formula is unchanged. However, at
the energy $6\times
\tentwenty$, $c\tau$ for the neutron is comparable to the interaction mfp, so
at higher energies the nucleon bending-angle is reduced by $\lsim 2$. }
\beq
\delta\theta\sim 0.5^\circ
\times \frac{Z\,B_{nG}}{E_{20}}\sqrt{\dmpc\,\lammpc}\,.
\eeq
\label{eq:bending}
Here $B_{nG}$ is the magnetic field in units of nanogauss, $E_{20}$ and $Z$ are
the particle energy in units of $10^{20}$~eV and charge, and the lengths $D$
and $\lambda$ are given in units of Mpc. It is thought likely that coherent
extragalactic fields are nanogauss in magnitude \cite{newKronberg}, in which
case super-GZK primaries from $\lsim 50$~Mpc will bend only a few degrees,
typically (but note that protons at $10^{19}$~eV will bend through $\sim
30^\circ$).
Thus, local models either postulate many invisible sources
isotropically-distributed with respect to the Galaxy to provide the roughly
isotropic flux observed above $\egzk$, or postulate a large extragalactic
magnetic field to isotropize over our Northern Hemisphere the highest-energy
particles\footnote {Some models postulate helium or iron nuclei as the
primaries, to increase the bending by the charge factors 2 and 26,
respectively. } from a small number of sources \cite{tempest}. A common
prediction for local models is little or no directional pairing on small
scales, especially when events with energy $\sim \egzk$ are included with the
$\tentwenty$~eV events\footnote {There is the possibility of pairing due to
magnetic focussing, if the projection of large-scale extragalactic magnetic
fields on our sky contains caustics \cite{caustics}, and if the incoherent
magnetic fields are sufficiently small. }. In particular, models invoking
randomly distributed, decaying super-massive particles (SMPs) as sources
\cite{SMPs}, and models invoking a large magnetic field with considerable
incoherent component, predict a chance distribution of observed events on the
sky.

   In the second category of models, the cosmic ray sources are put at
   cosmic distances ($\gsim 100$~Mpc, i.e. $z\gsim 0.02$).
These large-$z$ models mitigate the source issue.
However, the propagation problem is exacerbated due to
   the increased distance. Most proposals of this type postulate some stable,
   charge-neutral primary having limited interaction with the radiation
   background.
   Examples include the neutrino (which may regenerate a local nucleon/photon
   flux via ``Z-bursts'' \cite{Zburst},
or may develop a strong interaction at high energies \cite{strongNu},
   magnetic monopoles \cite{monopoles},
   and the lightest SUSY baryon \cite{susyB}
   (if the gluino mass is $\sim 1$~GeV).
   Other large-$z$ models employ broken Lorentz invariance
   or broken CPT-symmetry with effects generally suppressed by
   $M_P^{-1}$ factors but still large enough at $E\gsim \egzk$
   to suppress the nucleon-CMB interaction \cite{LIV}.
   The nucleon-magnetic field interaction
   may also be suppressed, in which case these nucleon primaries,
   like the charge-neutral primaries,
   are not significantly bent by the intervening extragalactic magnetic
   fields.
   A common prediction, then, of large-$z$ models is direct pointing of
   the primary's arrival direction back to the source\footnote
{There is some evidence that the highest energy events may indeed
   point to distant compact radio-loud quasars \cite{BF99};
however, more recent data do not seem to support the earlier result
\cite{STAR00}.
}.
   If a source is of sufficient intensity and duration,
   then the observation of multiple events from the direction of the source
   may be expected, beyond what is expected from chance coincidence.

So it is seen that a major discriminator between the local and the large-$z$
models is the occurrence of directional pairing compared to that expected from
chance coincidence alone. The AGASA experiment has already presented data
strongly suggesting that directional pairing is occurring at higher than chance
coincidence \cite{AGASApairs}. Comparisons of event directions in a combined
data sample of four experiments further supports non-chance coincidences
\cite{ALLpairs}. Furthermore, the energy-time correlation within pairs seems to
disfavor models with charged primaries originating from a common source of
relatively short duration.  This is because magnetic diffusion should lead to a
later arrival time (more bending) for the lower energy charged-primary in the
pair, contrary to what is observed in some pairs \cite{Et}.

It is the purpose of this paper to estimate in a straightforward manner the
significance of multiple events in future data. We do so by providing an
analytic calculation of the chance occurrence of directional multiples. It is
relative to these chance probabilities that observed rates will determine the
rise or fall of models. We present a formula for all possible multiples based
on statistics alone, as a function of the total number of observed events, and
the number of angular bins. We show that our analytic formula reproduces the
published AGASA probabilities (obtained by Monte Carlo simulation) fairly well,
even without inputting detailed knowledge of the experiment. We present the
chance occurrence of directional multiples for the HiRes experiment, and the
next-generation Auger and Telescope Array experiments, and finally the proposed
EUSO/OWL/AirWatch experiment. We show that for 20 events at HiRes, the
observation of a triplet or two doublets at resolution of 2\deg or less is
unlikely at the $3\sigma$ level, as is the observation of two triplets or a
quadruplet for 100 events at Auger. For the EUSO/OWL/AirWatch facility, the
anticipated very large number of events can be binned into a sizeable number of
subsets. As will be shown, this process can allow an explicit quantitative test
of the probabilities predicted for certain cluster groupings on the basis of
purely chance coincidence.


\section*{Formula for Chance Coincidence}

We now present the combinatoric formula for the probability
   of various event distributions in angle. Our formula is exact in the limit
where the experimental efficiency for observing events is effectively uniform
over the coverage of the celestial sky. We imagine that the sky coverage
consists
of a solid angle $\Omega$ divided into $N$ equal angular bins,
each with solid angle $\omega\simeq\pi\theta^2$~steradian;
the number of bins with cone half-angle $\theta$ is
   \beq
   N\simeq \frac{\Omega}{\pi\theta^2}
   = 1045\; \frac{(\Omega/1.0\ {\rm sr})}{(\theta/1.0^\circ)^2}
   \label{eq:bins}
   \eeq
   where $\Omega$ is the solid angle (sidereal or galactic)
on the celestial sphere covered by the experiment.
We toss $n$ events at random into these bins. As mentioned in the introduction,
such a chance distribution of events is just what is expected in some
models for the EECRs, e.g. randomly situated decaying SMPs, or charged-particle
or monopole primaries traversing incoherent magnetic fields.

Define each event distribution by specifying the partition of the $n$ total
events into a number $m_0$ of empty bins, a number $m_1$ of
single hits,a number $m_2$ double hits, etc.,
among the $N$ angular bins which constitute the total sky exposure.
The probability to obtain a given event topology is:
   \beq
   P(\{m_i\},n,N)=\frac{1}{N^n}\:
   \frac{N!}{m_0!\ m_1!\ m_2!\ m_3!\ldots}\:
   \frac {n!}{(0!)^{m_0}\ (1!)^{m_1}\ (2!)^{m_2}\ (3!)^{m_3}\ldots}\;.
   \label{eq:Hys}
   \eeq
The $N!$ and $n!$ factors in the numerator count the permutations of
the bins and the events, respectively.
The $m_j!$ and $j!$ factors in the denominator remove the
overcounting of those bins containing j events, and the events within
those bins, respectively.
The normalization factor $N^n$ is total number of ways
   to partition $n$ events among $N$ bins.

The variables in the probability are not all independent.
The partitioning of events is related to the total number of events by
   \beq
   \sum_{j=1} j\times m_j = n\,,
   \label{eq:sumjmj}
   \eeq
and to the total number of bins by
   \beq
   \sum_{j=0} m_j = N\,.
   \label{eq:summj}
   \eeq
Because of the constraints in eqs.\ (\ref{eq:sumjmj}) and (\ref{eq:summj}),
we infer that the process is not described by a simple multinomial
or Poisson probability distribution.

It is useful to use eqs.\ (\ref{eq:sumjmj}) and
(\ref{eq:summj}) to rewrite our exact probability (\ref{eq:Hys}) as
   \beq
   P(\{m_i\},n,N)=\frac{N!}{N^N}\,\frac{n!}{n^n}\,
   \prod_{j=0} \frac{(\ol{m_j})^{\,m_j}}{m_j!}\;,
  \label{eq:rewrite}
   \eeq
where we have defined
   \beq
   \ol{m_j}\equiv N\left(\frac{n}{N}\right)^j\frac{1}{j!}\;.
   \label{eq:meanm}
   \eeq
In the $n\ll N$~limit, $\ol{m_j}$
is expected to approximate the mean number of $j$-plets,
and eq. (\ref{eq:rewrite}) becomes roughly Poissonian.
As an approximate mean, $\ol{m_j}$ defined in eq.\ (\ref{eq:meanm})
provides a simple estimate of cluster probabilities due to chance
for the $n\ll N$~case.

Scaling laws relating mean cluster numbers to total event numbers and
binning angles may be derived by inserting eq.\ (\ref{eq:bins})
into eq.\ (\ref{eq:meanm}).
Results are
\beq
\ol{m_j}=\left(\frac{\pi n\theta^2}{\Omega}\right)^{j-1}
         \,\left(\frac{n}{j!}\right)\,,
\label{eq:meanscale}
\eeq
and
\beq
\frac{\ol{m_j}}{\ol{m_{j-1}}}=\frac{n}{jN}\sim \frac{\pi n\theta^2}{j\Omega}\,.
\label{eq:mratio}
\eeq

These scaling laws may be used as a further test of the
randomness of clustering in a data sample.
For example, the angular binning may be artificially expanded from the
experimental resolution to see
if the cluster numbers follow the $\theta^{2(j-1)}$~scaling law.
Of course, signal-to-chance is optimised by choosing the binning angle
close to the
natural angle of the source on the sky.

In the next section we examine the simplification
of the exact formulas (\ref{eq:Hys}) and (\ref{eq:rewrite})
that results in the large $N,n$ limit.
Even in the $N\gg n\gg1$ limit,
the resulting approximate formula will be seen to be not quite Poissonian,
since the variables in the set \{$m_i,n,N$\} are not independent.

\section*{Large $N, n$ Limit}
Two large-number limits of interest are $N\gg n\gg 1$, and $n>N\gg 1$.
With bin numbers typically $\sim 10^3$, the first limit applies to the
AGASA, HiRes, Auger and Telescope Array experiments; the second limit
becomes relevant for the EUSO/OWL/AW experiment after a year or more of running.

\subsection*{Approximation for $N\gg n\gg 1$}
When $N\gg n$, the number $m_0$ of empty bins is of order $N$,
and the number of bins $m_1$ with single events (singlets) is order $n$;
the number of clusters (doublets, triplets, etc.) is small.
It is sensible to explicitly evaluate the not-so-interesting
$j=0$~and~1 terms in
eqs.\ (\ref{eq:rewrite}) and (\ref{eq:meanm}).  With the use of Stirling's
approximation for the factorials, one arrives at
%
%
a simple form for the probability, valid when $N\gg n\gg 1$:
   \beq
P(\{m_i\},n,N)\approx {\cal P}
\left[ \prod_{j=2} \frac{(\ol{m_j})^{m_j}}{m_j!}
e^{-\ol{m_j}\,r^j (j-2)!} \right]\,,
\label{eq:largeN}
   \eeq
where $r\equiv (N-m_0)/n\approx 1$, and the prefactor ${\cal P}$ is
\beq
{\cal P}=e^{-(n-m_1)}\,\left(\frac{n}{m_1}\right)^{m_1 +\half}\,.
\eeq
In the ``sparse events'' case here,
where $N\gg n$, one expects the number of singlets $m_1$
to approximate the number of events $n$.  In this case the prefactor
is near unity.
The non-Poisson nature of Eq.\ (\ref{eq:largeN}) is reflected in the
factorials and powers of $r$ in the exponents,
and the deviation of the prefactor from unity.
In our numerical work, we will provide curves for the exact result,
and for the Poisson approximation (obtained from our
expressions (\ref{eq:largeN}) and (\ref{eq:meanm}) by setting ${\cal P}$
to unity and omitting $r^j\,(j-2)!$ in the exponent, and by
neglecting the constraints of eqs.\ (\ref{eq:sumjmj}) and (\ref{eq:summj}).
Note that in this Poisson approximation, $\ol{m_j}$ as defined in
Eq.\ (\ref{eq:meanm}) is truly the mean number of $j$-plets.

\subsection*{Approximation for $n>N\gg 1$}
In the case where $n>N\gg 1$, higher $j$-plets are common and
the distribution of clusters can be rather broad in $j$,
since according to Eq.\ (\ref{eq:mratio}).

Already at $j=1\,(2)$, Stirling's approximation to $j!$ is good to $8\%
\,(4\%)$,
and so we may write $\ol{m_j}$ in the approximate form for $j\ge 1$:
\beq
\ol{m_j}\approx \sqrt{\frac{N^3}{2\pi
en}}\,\left(\frac{en}{jN}\right)^{j+\half}\,.
\label{eq:appm}
\eeq
Extremizing this expression with respect to $j$,
one learns that the most populated $j$-plet occurs near $j\sim n/N$.
Combining this result with the broad distribution expected for large $n/N$,
one expects clusters with $j$ up to ${\rm several}\times \frac{n}{N}$
to be common in the EUSO/OWL/AW experiment.


   \section*{Comparison with Experiments}
   There are two ongoing EECR experiments, AGASA \cite{AGASA}
and HiRes \cite{HiRes}.
  There are two larger area experiments under development,
   Auger \cite{Auger} and Telescope Array (TA) \cite{TA}.
   Finally, there are experiments proposed with still larger areas,
   EUSO \cite{EUSO}, OWL \cite{OWL} and AirWatch \cite{AW}.
   A non-hostile merger of these latter experiments appears likely,
   so we refer to them collectively as EUSO/OWL/AW.
   In Table 1 we list the relevant parameters defining for our purposes
   each of these experiments.

   \begin{table}
   \caption[]
{Typical values of effective area $A$,
celestial solid angle $\Omega$ \cite{Omega}, and angular resolution
$\theta_{\rm min}$ for the existing and proposed EECR experiments.
The incident flux $F(\ge \egzk)=10^{-19}{\rm cm^{-2} s^{-1} sr^{-1}}$
has been used to estimate the number of events above $\egzk=5\times
10^{19}$~eV.}
   \vspace{0.5 cm}
   \centering\leavevmode
   \begin{tabular}{|c|c|c|c|c|}
   \hline
  Experiment & AGASA & HiRes & Auger/TA & EUSO/OWL/AW \\ \hline
   $A\: ({\rm km^2\,sr})$ & 150 & 800 & $6\times 10^3$ & $3\times 10^5$ \\
   \hline
   $n/{\rm yr}=A\times F(\ge \egzk)$ & 5 & 30 & 200 & $ 10^4$ \\
\hline
$\Omega\: ({\rm sr})$ & 4.8 & 7.3 & 4.8 & 4$\pi$ \\ \hline
   $\theta_{\rm min}$ & $3.0^\circ$ & $0.5^\circ$ & $ 1.0^\circ$ &
$1.0^\circ$ \\ \hline
   \end{tabular}
   \end{table}

We proceed to normalize the analytic approach described above against the
Monte Carlo method used by the AGASA Collaboration \cite{AGASApairs},
and then to provide some concrete
   examples for future observation.
   In searching for cluster probabilities using Monte Carlo, a fixed number
   $n$ of
   events is tossed into phase space, and clusters are defined by choosing a
   correlation angle $\theta\pri$ (in principle independent from the
resolution angle
discussed earlier). For example,
   a doublet is registered when an event
   falls within an angle $\theta\pri$ of a preceding event. In such a manner,
   the AGASA Collaboration \cite{AGASApairs}
found that 92 events yielded 12 or
   more doublets in 1.5\% of their trials. Their simulations utilized a
   correlation angle $\theta\pri=3\deg$ and declination angles (roughly)
   between
   10\deg and 70\deg . This gives a solid angle $\Omega\simeq
4.8$~steradian. If for
   the
   moment we identify the correlation angle $\theta\pri$ with
   the resolution angle $\theta,$ then from
   \eq{bins}, with $\theta=3\deg,$ we find $N=557.$ Eq.(\ref{eq:Hys}) then
   gives
   \be
   P(m_2\ge 12,\ m_3\ge 0,\ m_4\ge 0,\ 92,\ 557)=1.4\mbox{\%}\ \ .
  \labeq{j}
   \ee
Considering the crudeness of the approximations, this result agrees
   with the 1.5\%
   simulation result.\footnote
{Note that if our estimate of the solid angle
   were to
   change somewhat, agreement with the Monte Carlo results could still be
   achieved by introducing a slight deviation from the assumed equality
   $\theta\pri=\theta.$}

   For Auger, we adopt the same solid angle $(\Omega=4.8\ {\rm sr}).$ For
HiRes, we
   estimate on geometric grounds a solid angle of 10.9 sr \cite{Omega}.
The same geometric
   estimate for AGASA gives a solid angle of 6.9 sr, which the acceptance
profile
   reduced to the above-mentioned 4.8 sr. For the sake of our modeling, we
   apply the same reduction factor $(4.8/6.9)$ to the HiRes geometry, to arrive
   at an effective solid angle of 7.3 sr (Table I).
   The purely statistical probabilities for various cluster topologies can now
   be
   calculated as a function of the angular resolution and the accumulated
   number
   $n$ of events.

   \section*{Discussion of Results}
\subsection*{HiRes}
   For the HiRes experiment, about 20 events at $10^{20}$~eV
are expected when the first full year's data is analyzed.
We calculate the inclusive probabilities for
one or more, two or more, and three or more doublets;
   and one or more triplets, over a range of angular
   binning using Eqs.~\req{Hys}and \req{bins}.
Note that by ``inclusive probability'' we mean the stated
number of $j$-plets {\it plus any other clusters};
e.g. a topology with two doublets and one triplet counts as
one doublet, as two doublets, and as one triplet in the inclusive probabilities.
The results are displayed in Figure 1.
%
%
\begin{center}
\includegraphics[scale=.7]{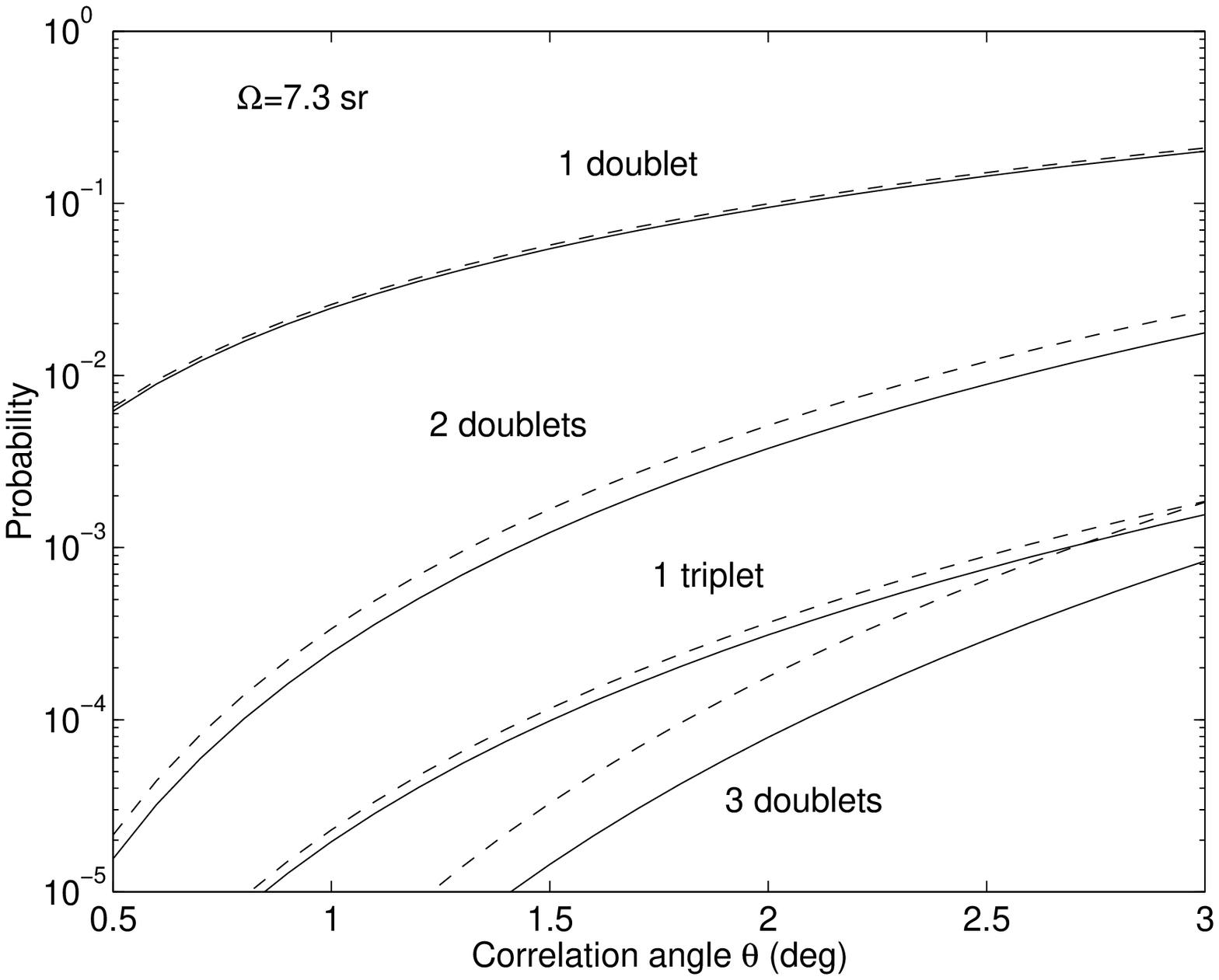}
{\renewcommand{\baselinestretch}{1.2}
\hspace*{.1in}\parbox[t]{4.5in}
{\small Figure 1: Inclusive probabilities for various clusters, given 20
events at HiRes.
The solid line is the exact result, the dashed line is the Poisson
approximation.}}
   \end{center}
\bigskip

\noindent Several comments may be made with reference to this figure:
   \bi
   \itm{a} For all except the 3 doublet configuration, the
   Poisson approximation using the mean values in Eq.(\ref{eq:meanm})
provides an estimate good to
   within 50\% of the non-approximate form;
for the (much suppressed) 3 doublet configuration,
it overestimates
the probability by about a factor of 3 in much of the angular region.
   \itm{b} For angular binning tighter than 2\deg, an observation of two
   doublets among the first 20 events has a chance probability of less than
0.5\%.
   Thus the observation of this topology could be construed
   as evidence (at the 3$\sigma$ level) for clustering beyond statistical. The
   observation of a triplet within $\le 3\deg$ has a random probability of less
   than $10^{-3},$ and hence observation of such a triplet would most
likely signify
   clustered or repeating sources, or magnetic focussing effects.
  \itm{c} With the accumulation of 40 events (not shown in the figure),
the appearance of two
   doublets
   has a probability of less than 0.5\% for a correlation angle
   of 1\deg or less. This illustrates how the good
   angular resolution of HiRes may be used to detect non-statistical
   clustering
   with only a few observed clusters.\ei
\subsection*{Auger}

Coming now to Auger, we  present in Figure 2 the probabilities of
observing 8 doublets, one triplet, and two  triplets,
in an event sample of 100 events.
   \begin{center}
\includegraphics[scale=0.7]{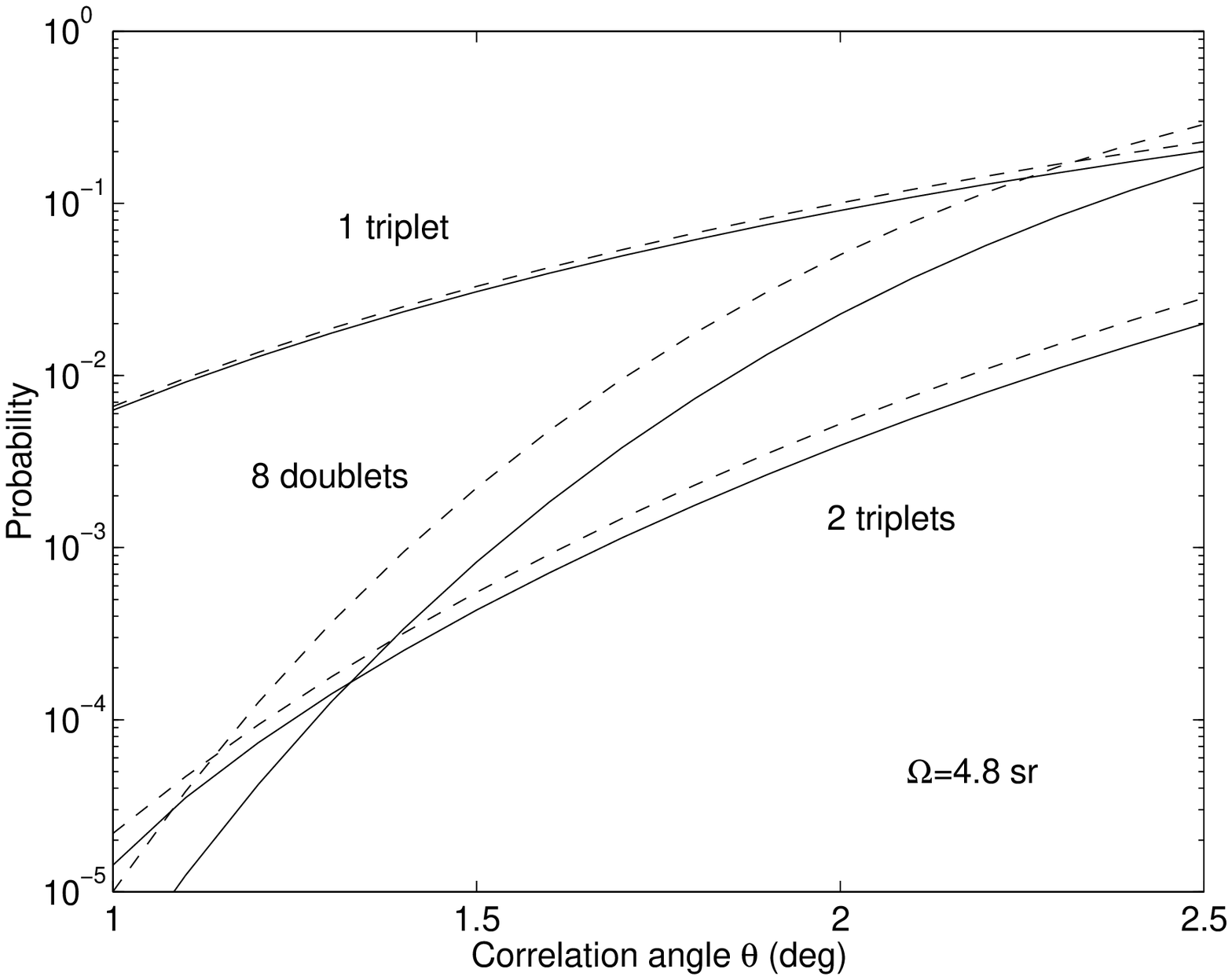}
\nopagebreak
{\renewcommand{\baselinestretch}{1.2}
\hspace*{.1in}\parbox[t]{4.5in}
{\small Figure 2: Inclusive probabilities for various clusters in a 100
event sample at Auger;
solid (exact), dashed (Poisson).}
}

   \end{center}
\bigskip

This graph illustrate several points of interest:
\bi
\itm{a} The 8-doublet probability is extremely sensitive to the angular
binning, and thus
uncertainties of the order of 0.5\deg in the region of $\theta\le 2\deg$ would
preclude assigning a baseline chance-probability to better than an order of
magnitude
for this topology.
This uncertainty may be avoided by breaking down the 100 events into
smaller data sets.
On the other hand, it may be possible to use the sensitive dependence
displayed here to advantage:
observation of a flatter dependence on angular bin-size could signal a
non-chance
origin for the clustering.
\itm{b} As in Fig. 1, it is seen that the Poisson approximation
is good for some topologies, but an overestimate for others;
for 8 doublets, it overestimates
the probability by about 3 in much of the angular region.
\itm{c}
The observation of two triplets with angular binning of less than 2\deg would
have a random probability of less than 0.5\%, and hence could be construed as
3-sigma evidence for a novel astrophysics, such as
clustered or repeating sources or magnetic focussing.
\itm{d}Not shown in the
figure is the probability of observing a quadruplet, which turns out to be
about 0.5\% for a binning angle of 2.5\deg. Hence the observation of a
quadruplet within 2.5\deg\ among the first 100 Auger events would be suggestive
of clustering due to an astrophysical cause.
\ei

\subsection*{EUSO/OWL/AW}
The large number of events expected in the EUSO/OWL/AW experiment
presents an extraordinary opportunity to test for non-random
clustering, but also a numerical problem for both our formula
and for a direct Monte Carlo simulation.
With $10^4$ super-GZK events (Table I) disributed among $10^3$ bins,
10-plets and beyond may be common ocurrences (see eq.\ (\ref{eq:appm})).
One approach to the very large data sample
is to envisage the $10^4$ super-GZK
events divided into 500 more manageable subsamples of 20 events each
(there are many ways to choose the event partitions,
and some subtleties are involved.).
Then the probability predictions for clustering in the random model
can be straightforwardly assessed.
For example, In Figure 3, we show the inclusive probabilities for
1 and 2 doublets in 20 events, for the EUSO/OWL/AW sky aperture.
\begin{center}
\includegraphics[scale=0.7]{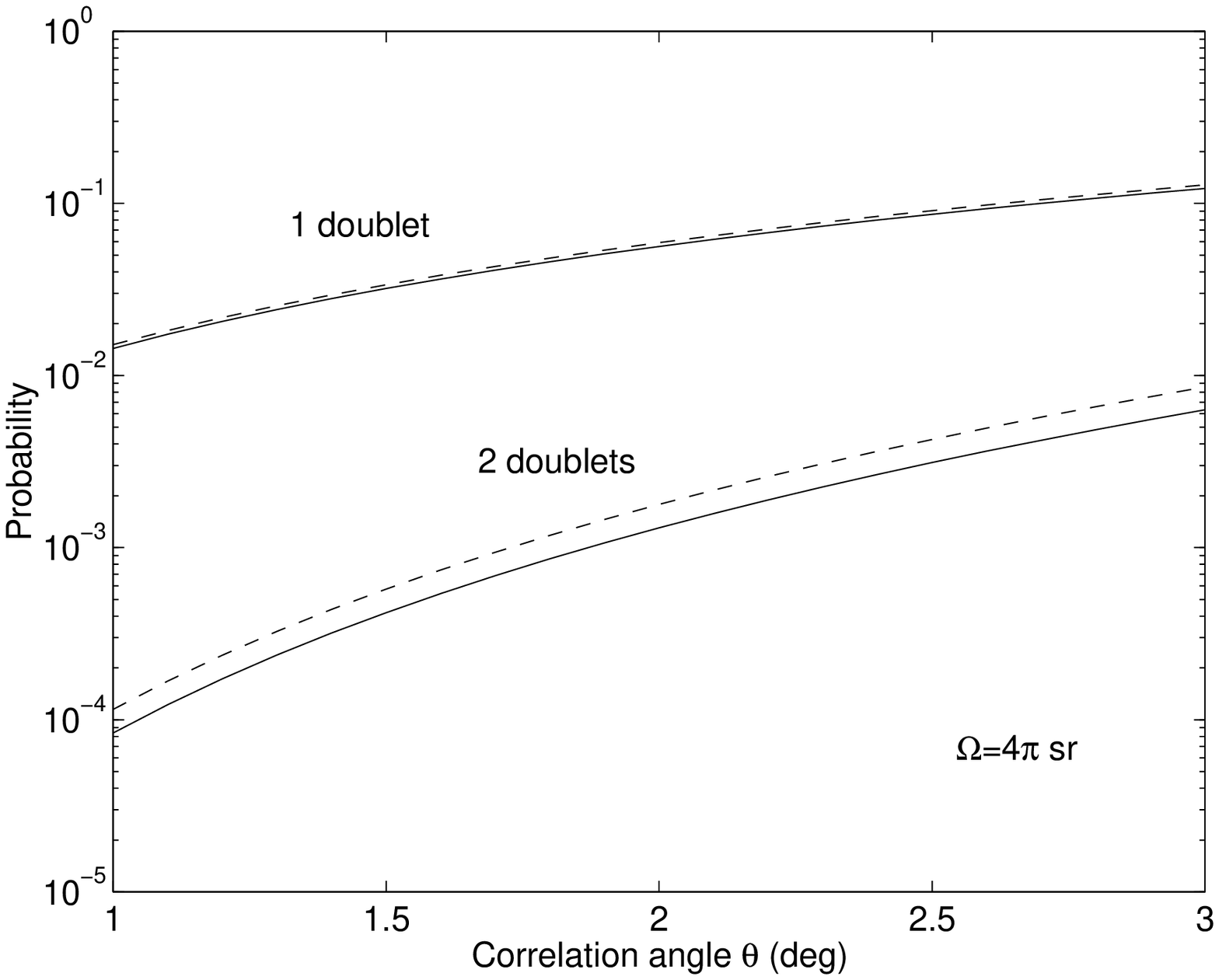}
\nopagebreak
{\renewcommand{\baselinestretch}{1.2}
\hspace*{.1in}\parbox[t]{4.5in}
{\small Figure 3: Probabilities for various topologies in a 20 event
subsample at EUSO/OWL/AW}\bigskip

 }  
   \end{center}

We can see that for binning angles of  2.5\deg, we expect about 10\% of the
500 20-event subsamples, or about 50 subsamples to have one doublet, and we
expect
about 0.3\%, or perhaps 1 or 2 samples to have 2 doublets.
Comparing subsamples in this way, deviations
from random clustering can be {\em quantitatively}  assessed.
Of course, it is possible that in the large sample of EUSO/OWL/AW,
some non-random high-$j$ clusters will emerge far above background.
In such a case, the random probabilites presented here become much less
relevant.

The same large data-set approach just described could also be used
for Auger after a few years of running time,
although with somewhat fewer statistics.

\section*{Summary and Concluding Remarks}
%
We have presented an analytic study of clustering for
cosmic rays based on a random angular generation of events in the
sky. Our probability formula is based on randomly assigning events
into fixed angular bins with uniform a priori probability.
In reality, the efficiency of experiments for observing events is not uniform
across the sky coverage.
For this reason, the most careful quantitative statements about clustering
probabilities of existing data must come from Monte Carlo simulations
incorporating experimental efficiencies.
Nevertheless, our results are in good agreement with the
prior Monte Carlo study by the AGASA Collaboration \cite{AGASApairs}.
For this reason, we believe that our formula makes a significant
advance to the field, and is especially useful in predicting event topologies
for future experiments and larger data samples.

We found that the use of Poisson distributions, with mean values given
by \eq{meanm}, was approximately valid for some topologies, but
yielded overestimates for others.
For some of the interesting cases discussed in the preceding section, the
Poisson estimates were factors of 2 to 3 larger than the exact probabilities.

Results for the HiRes, Auger and EUSO/OWL/AW experiments were
presented. These results reveal which topologies occur
with probabilities of 0.5\% or less in the various experiments;
observation of these topologies would constitute evidence at the
$3\sigma$ level for astrophysical rather than random causes for the clustering.
An observation of two or more
doublets in the first 20 events at HiRes, each doublet consisting
of 2 events within less than 2\deg of each other, is one example shown
in the text.

Topologies with highly suppressed chance probabilities are especially sensitive
probes of non-random clustering. This situation is exemplified in our
discussion of the 8-doublet topology for the Auger experiment with 100 events.
Highly suppressed topologies may be rather difficult to use in practice since
they exhibit extreme sensitivity to binning angle, which leads to great
ambiguity in their statistical significance. On the other hand, this
sensitivity may be useful as a diagnostic to distinguish between random
clustering vs.\ clustering due to astrophysics.

The large number of events previewed in the
proposed EUSO/OWL/AW experiment (and to a lesser extent, at the Auger
facility) presents an analytical challenge.
Partitioning of the total data sample into subsamples,
and then comparing these subsamples, would provide a direct test
of the purely statistical predictions for clustering.

We have not included any source modeling in our analysis.
Our chance probabilities describe arrival directions
randomly distributed on the celestial sky.
In fact, this distribution is reality for some models,
such as decaying SMPs, and charged primaries with directions
randomized by incoherent cosmic magnetic fields.
A complementary approach to our work is to consider specific source models
generating non-random angular distributions. Steps along
this line of inquiry have recently been taken \cite{nonrandom}.
Future progress in the field will involve comparisons of the
random and non-random model predictions with the data.


\section*{Acknowledgements}

We acknowledge fruitful discussions with Jim Fry,
   and thank the Aspen Center for Physics for a beneficial working environment.
   This work was supported in part by the
   U.S. Department of Energy grant no.\ DE-FG05-85ER40226, and by the National
   Science Foundation grant no.\ PHY-9722044.


\begin{thebibliography}{99}
\bibitem{Omega}
To estimate the celestial sky coverage $\Omega$
for ground-based experiments,
one rotates the fixed-coordinate solid angle about the earth's axis of rotation.
For experiments with vertical acceptance from the zenith to $\theta_z$,
e.g. AGASA and Auger/TA, the
relevant formula is
%
\[
\Omega = 2\pi\left\{
\begin{array} {ll}
2\sin\theta_z\cos\alpha,\quad & {\rm for}\quad \alpha+\theta_z < 90^\circ \\
     1+\sin(\theta_z -\alpha),\quad & {\rm for}\quad \alpha+\theta_z > 90^\circ
\end{array}
\right.
\]
%
where $\alpha$ is the latitude of the experiment,
which is $36^\circ$N for AGASA and $35^\circ$S for Auger.

For experiments with horizontal acceptance from $\theta_1$
to $\theta_2$ with respect to the horizon, e.g. HiRes,
%
\[
\Omega=2\pi \left\{
\begin{array}{ll}
   \cos(\alpha+\theta_1)+\cos(\alpha-\theta_2),\quad
   & {\rm for}\quad \theta_2 < \alpha  \\
\cos(\alpha+\theta_1) + 1,\quad & {\rm for}\quad \theta_2 >\alpha
\end{array}
\right.
\]
%
For HiRes, $\alpha \sim 40^\circ$N.
Finally, for an orbiting experiment such as EUSO/OWL/AW,
the celestial coverage is nearly $4\pi$ steradians.
\end{thebibliography}
\end{document}